\documentstyle[epsfig,12pt]{article}
\newcommand{\zp}[3]{Z. Phys.\ C#1 (19#2) #3}
\newcommand{\pl}[3]{Phys.\ Lett.\ #1B (19#2) #3}
\newcommand{\np}[3]{Nucl.\ Phys.\ B#1 (19#2) #3}
\newcommand{\prd}[3]{Phys.\ Rev.\ D#1 (19#2) #3}
\newcommand{\prl}[3]{Phys.\ Rev.\ Lett.\ #1 (19#2) #3}

\newcommand{\md}{\mbox{d}}
\def\simgt{\rlap{\lower 3.5 pt \hbox{$\mathchar \sim$}} \raise 1pt \hbox {$>$}}
\def\simlt{\rlap{\lower 3.5 pt \hbox{$\mathchar \sim$}} \raise 1pt \hbox {$<$}}

\newcommand{\beq}{\begin{equation}}
\newcommand{\eeq}{\end{equation}}
\newcommand{\bea}{\begin{eqnarray}}
\newcommand{\eea}{\end{eqnarray}}

\catcode`\@=11

\def\@citex[#1]#2{\if@filesw\immediate\write\@auxout{\string\citation{#2}}\fi
  \def\@citea{}\@cite{\@for\@citeb:=#2\do
    {\@citea\def\@citea{,\penalty\@m}\@ifundefined
       {b@\@citeb}{{\bf ?}\@warning
       {Citation `\@citeb' on page \thepage \space undefined}}%
\hbox{\csname b@\@citeb\endcsname}}}{#1}}

\def\citer{\@ifnextchar [{\@tempswatrue\@citexr}{\@tempswafalse\@citexr[]}}

\def\@citexr[#1]#2{\if@filesw\immediate\write\@auxout{\string\citation{#2}}\fi
  \def\@citea{}\@cite{\@for\@citeb:=#2\do
    {\@citea\def\@citea{--\penalty\@m}\@ifundefined
       {b@\@citeb}{{\bf ?}\@warning
       {Citation `\@citeb' on page \thepage \space undefined}}%
\hbox{\csname b@\@citeb\endcsname}}}{#1}}
\relax

\begin{document}

\thispagestyle{empty}

\hfill\vbox{\hbox{\bf DESY 95-046}
                                   }
\vspace{1.5in}

\begin{center}
\boldmath
{\large\bf INELASTIC $J/\psi$ PHOTOPRODUCTION$\,^*$} \\
\unboldmath

\vspace{1.5cm}

{\large \sc Michael~Kr\"amer}$\,^\dagger$

\vspace{0.3cm}

{\em Deutsches Elektronen-Synchrotron DESY, D-22603 Hamburg, FRG}

\end{center}

\vspace{1.5cm}

\noindent
\begin{abstract}
Inelastic photoproduction of $J/\psi$ particles at high energies
is one of the processes to determine the gluon distribution in the nucleon.
The QCD radiative corrections to the color-singlet model of this reaction
have recently been calculated. They are large at moderate photon energies,
but decrease with increasing energies. I compare the cross section and the
${J/\psi}$ energy spectrum with the available fixed-target photoproduction
data. Predictions for the HERA energy range are given which
demonstrate the sensitivity of the result to the parametrization of the
gluon distribution in the small-$x$ region.
\end{abstract}

\vfill

\noindent
{\footnotesize
$*\,$ Talk presented at the Workshop on "Heavy Quark Physics",
Bad Honnef, FRG, Dec. 1994.\\
$\dagger\,$ E-mail: mkraemer@desy.de}
\newpage

\noindent
The measurement of the gluon distribution in the nucleon is one of the
important goals of lepton-nucleon scattering experiments. The classical
methods exploit the evolution of the nucleon structure functions with
the momentum transfer and the size of the longitudinal structure function.
With rising energies, however, jet physics and the production of heavy
quark states become important complementary tools. Besides open charm and
bottom production, the formation of $J/\psi$ bound states \cite{bj81} in
inelastic photoproduction experiments
\begin{equation}
\gamma + {\cal N} \to J/\psi + X
\end{equation}
provides an experimentally attractive method since $J/\psi$ particles are
easy to tag in the leptonic decay modes.

Many channels contribute to the generation of $J/\psi$ particles in
photoproduction experiments \cite{jst92}, similar to hadroproduction
experiments. However, no satisfactory quantitative picture has emerged
yet and the production of a large surplus of $\psi'$ particles in
$p\overline{p}$ collisions awaits the proper understanding. Theoretical
interest so far has focussed on two mechanisms for $J/\psi$ photo- and
electroproduction, elastic/diffractive \cite{elastic,elastexp} and
inelastic production through photon-gluon-fusion \cite{bj81,jst92}.
While by the first mechanism one expects to shed light on the physical
nature of the pomeron, inelastic $J/\psi$ production provides information
on the distribution of gluons in the nucleon \cite{mns87}. The two
mechanisms can be separated by measuring the $J/\psi$ energy spectrum,
described by the scaling variable
$z = {p\cdot k_\psi}\, / \, {p\cdot k_\gamma}$
with $p, k_{\psi,\gamma}$ being the momenta of the nucleon and $J/\psi$,
$\gamma$ particles, respectively. In the nucleon rest frame, $z$ is
the ratio of the $J/\psi$ to the $\gamma$ energy, $z=E_\psi/E_\gamma$.
For elastic/diffractive events $z$ is close to one; a clean sample of
inelastic events can be obtained in the range $z<0.9$ \cite{brug}.
The production of $J/\psi$ particles at large transverse momenta is
dominated by gluon fragmentation mechanisms \cite{bdfm94}.
Additional production mechanisms, such as $B\overline{B}$ production
and the "resolved photon" contributions at HERA can be strongly suppressed
by suitable cuts \cite{jst92}.

\begin{figure}[hbtp]

\vspace*{-0.5cm}
\hspace*{0.75cm}
\epsfig{%
file=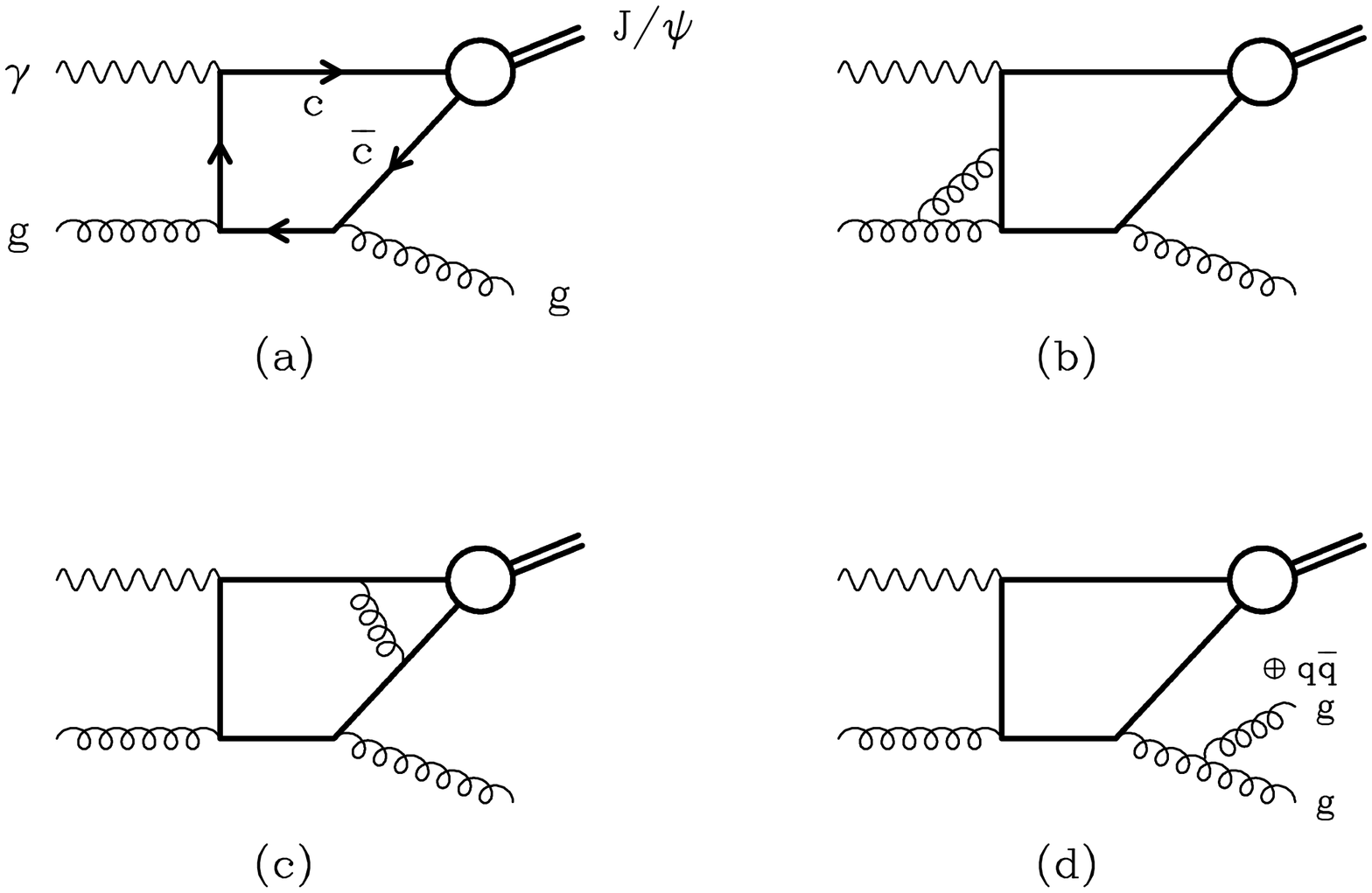,%
height=10cm,%
width=14.cm,%
bbllx=1.0cm,%
bblly=1.9cm,%
bburx=19.4cm,%
bbury=26.7cm,%
rheight=8.2cm,%
rwidth=15cm,%
angle=-90}


\hspace*{0.75cm}
\epsfig{%
file=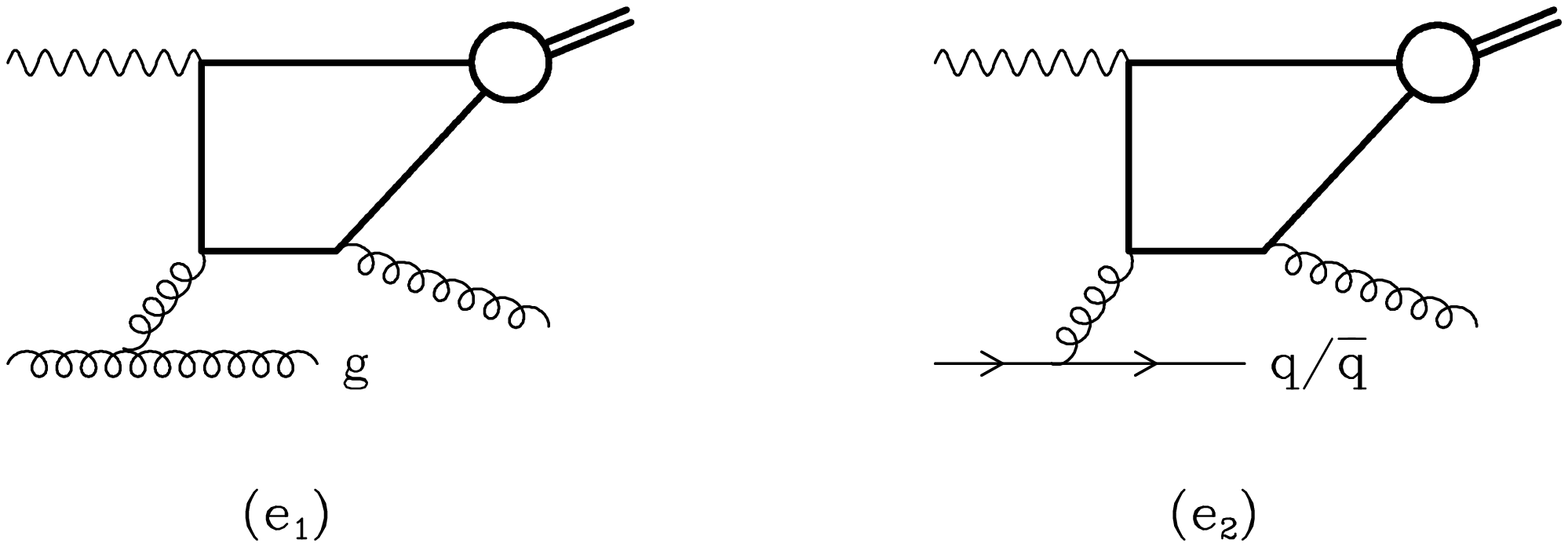,%
height=10cm,%
width=14.cm,%
bbllx=1.0cm,%
bblly=1.9cm,%
bburx=19.4cm,%
bbury=26.7cm,%
rheight=8.2cm,%
rwidth=15cm,%
angle=-90}

\vspace*{-3.2cm}

\caption[xx]{  \label{f_diagrams}
Generic diagrams for inelastic $J/\psi$ photoproduction:
(a) leading order contribution;
(b) vertex corrections;
(c) box diagrams;
(d) splitting of the final state gluon into gluon or
    light quark-antiquark pairs;
(e) diagrams renormalizing the initial-state parton densities.}

\vspace*{-5mm}

\end{figure}

Inelastic $J/\psi$ photoproduction through photon-gluon fusion is described
in the color-singlet model through the subprocess
\begin{equation}\label{eq_subpr}
\gamma + g \to J/\psi + g
\end{equation}
shown in Fig.\ref{f_diagrams}.
Color conservation and the Landau-Yang theorem require the emission of a
gluon in the final state. The cross section is generally calculated in the
static approximation in which the motion of the charm quarks in the bound
state is neglected. In this approximation the production amplitude factorizes
into the short distance amplitude $\gamma + g \to c\overline{c} + g$, with
$c\overline{c}$ in the color-singlet state and zero relative velocity of the
quarks, and the $c\overline{c}$ wave function $\varphi(0)$ of the $J/\psi$
bound state at the origin which is related to the leptonic width.
When confronted with photoproduction data of fixed-target
experiments \cite{na14,ftps}, the theoretical predictions underestimate
the measured cross section in general by more than a factor two, depending
in detail on the $J/\psi$ energy and the choice of the parameters \cite{jst92}.
The discrepancy with cross sections extrapolated from electroproduction
data \cite{emc,nmc} is even larger.

The lowest-order approach to the color-singlet model demands several
theoretical refinements: (i) Higher-order perturbative QCD corrections;
(ii) Relativistic corrections due to the motion of the charm quarks in the
$J/\psi$ bound state; and last but not least, (iii) Higher-twist effects
which are not strongly suppressed due to the fairly low charm-quark mass.
While the relativistic corrections have been demonstrated to be under
control in the inelastic region \cite{jkgw93}, the problem of higher-twist
contributions has not been approached so far. The calculation of the
higher-order perturbative QCD corrections has been performed
recently \cite{phd}. Expected {\em a priori} and verified subsequently, these
corrections dominate the relativistic corrections in the inelastic
region, being of the order of several $\alpha_s(M_{J/\psi}^{2})\sim 0.3$.
In the first step of a systematic expansion, they can therefore be
determined in the static approach \cite{bbl94}.

A detailed analysis of the ${\cal O}(\alpha\alpha_s^3)$ corrections to
inelastic $J/\psi$ photoproduction is the subject of a forthcoming publication
\cite{mk95}; first results have been presented in Ref.\cite{kzsz94}.
In this short note the implications of the higher order QCD corrections
for the partonic cross sections are discussed and the ${J/\psi}$ energy
spectrum is compared with the available fixed-target photoproduction data.
In addition, predictions for the HERA energy range are given which
demonstrate the sensitivity of the result to the parametrization of the
gluon distribution in the small-$x$ region.

Generic diagrams which build up the cross section in next-to-leading order
are depicted in Fig.\ref{f_diagrams}.
Besides the usual self-energy diagrams
and vertex corrections for photon and gluons (b), one encounters box
diagrams (c), the splitting of the final-state gluon into gluon and light
quark-antiquark pairs, as well as diagrams renormalizing the initial-state
parton densities (e). The evaluation of these amplitudes has been performed
in the Feynman gauge and the dimensional regularization scheme has been
adopted to calculate the singular parts of the amplitudes. The masses of
light quarks in Fig.\ref{f_diagrams}(d,e$_1$) have been neglected while
the mass parameter of the charm quark has been defined on-shell.
I have carried out the renormalization program in the extended
$\overline{\mbox{MS}}$ scheme \cite{msbarext} in which the massive particles
are decoupled smoothly for momenta smaller than the quark mass.
The exchange of Coulombic gluon quanta in the diagram (1c) leads to a Coulomb
singularity $\sim \pi^2/2\beta_R$ which can be isolated by introducing a
small relative quark velocity $\beta_R$. Following the standard path
\cite{hb57}, this effect has to be interpreted as the Sommerfeld rescattering
correction which can effectively be mapped into the $c\overline{c}$ wave
function. As expected, the infrared singularities cancel when the emission
of soft and collinear final-state gluons and light quarks, characterized by a
cut-off $\Delta$ \cite{mk95,bkns89,sn92}, is added to the virtual corrections.
The collinear initial-state singularities can be absorbed, as usual, into the
renormalization of the parton densities \cite{aem79} defined in the
$\overline{\mbox{MS}}$ factorization scheme.

The perturbative expansion of the photon-parton cross section can be
expressed in terms of scaling functions,
\beq
\hat\sigma_{i\gamma}(s,m_c^2) =
\frac{\alpha\alpha_s^2 e^2_c}{m_c^2}\,\frac{|\varphi(0)|^2}{m_c^3}
\left[ c_{i\gamma}^{(0)}(\eta) + 4\pi\alpha_s \left\{c_{i\gamma}^{(1)}(\eta)
+\overline{c}^{(1)}_{i\gamma}(\eta)\ln\frac{Q^2}{m_c^2}\right\}\right]
\eeq
$i=g,q,\overline{q}$ denoting the parton targets. For the sake of simplicity,
I have identified the renormalization scale with the factorization scale
$\mu_R^2 = \mu_F^2 = Q^2$. The scaling functions depend
on the energy variable $\eta = s/4m_c^2 - 1$. $c_{\gamma g}^{(0)}$ is the
lowest-order
contribution which scales $\sim \eta^{-1}\sim 4m_c^2/s$ asymptotically.
$c_{\gamma g}^{(1)}$ can be decomposed into a "virtual + soft" (V+S) piece
and a "hard" (H) gluon-radiation piece. The $\ln^j\Delta$ singularities of
the (V+S) cross section are mapped into (H), cancelling the equivalent
logarithms in this contribution so that the limit $\Delta\to 0$ can safely be
carried out.
The nomenclature "hard" and "virtual + soft" is therefore a
matter of definition, and negative values of $c^{(\mbox{\scriptsize{}H})}$
may occur in some regions of the parameter space. Up to this order, the
wave-function at the origin is related to the leptonic $J/\psi$ width by
\beq\label{eq_decaycorr}
\Gamma_{ee} = \left(1 - \frac{16}{3}\frac{\alpha_s}{\pi}\right)
\frac{16\pi\alpha^2e_c^2}{M_{J/\psi}^{2}}\, |\varphi(0)|^2
\eeq
with only transverse gluon corrections taken into account explicitly
\cite{lepdecay}.

\begin{figure}[hbtp]

\vspace*{-0.7cm}
\hspace*{0.75cm}
\epsfig{%
file=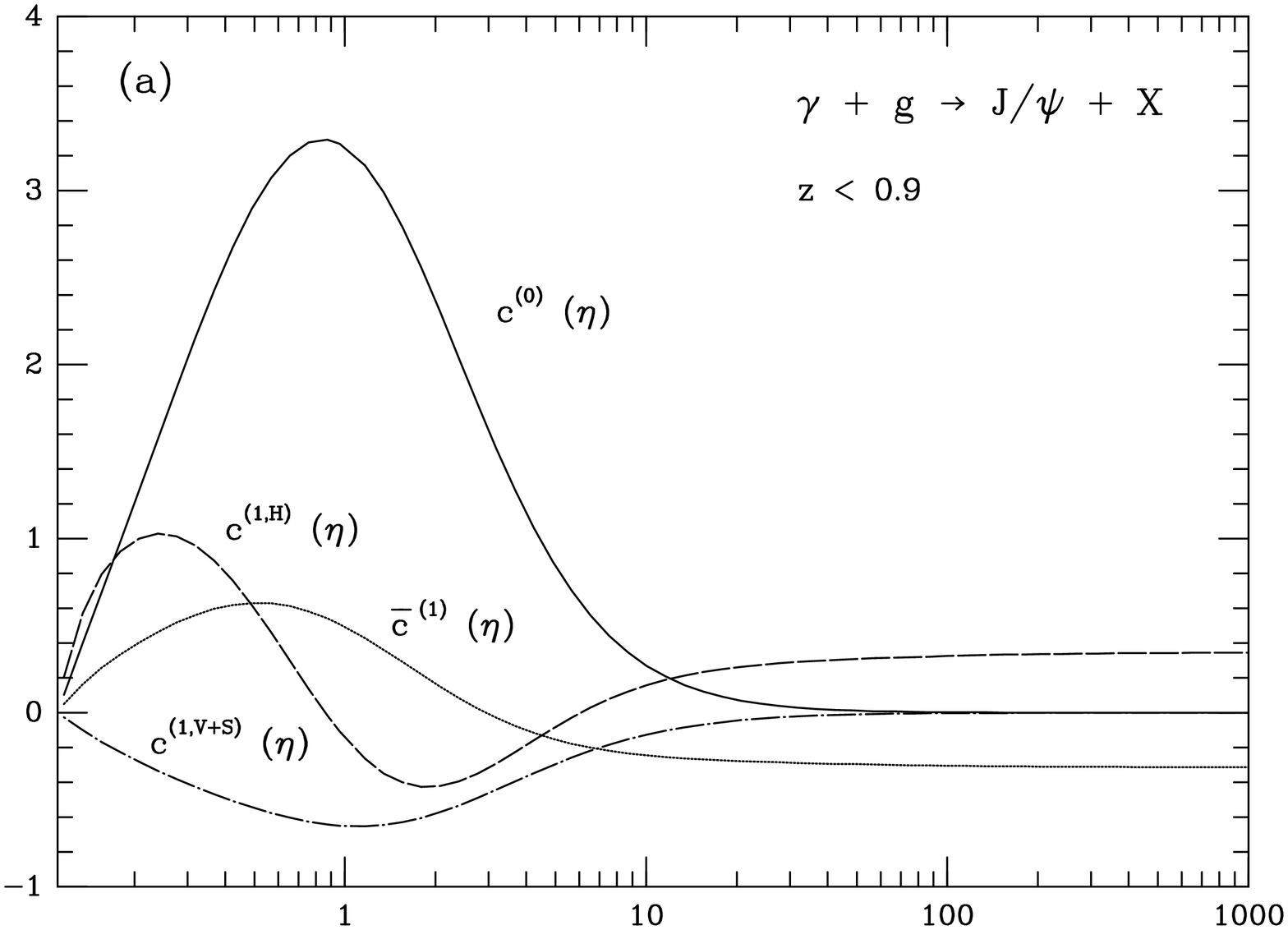,%
height=10cm,%
width=14.cm,%
bbllx=1.0cm,%
bblly=1.9cm,%
bburx=19.4cm,%
bbury=26.7cm,%
rheight=8.2cm,%
rwidth=15cm,%
angle=-90}

\vspace*{2.3cm}

\hspace*{0.75cm}
\epsfig{%
file=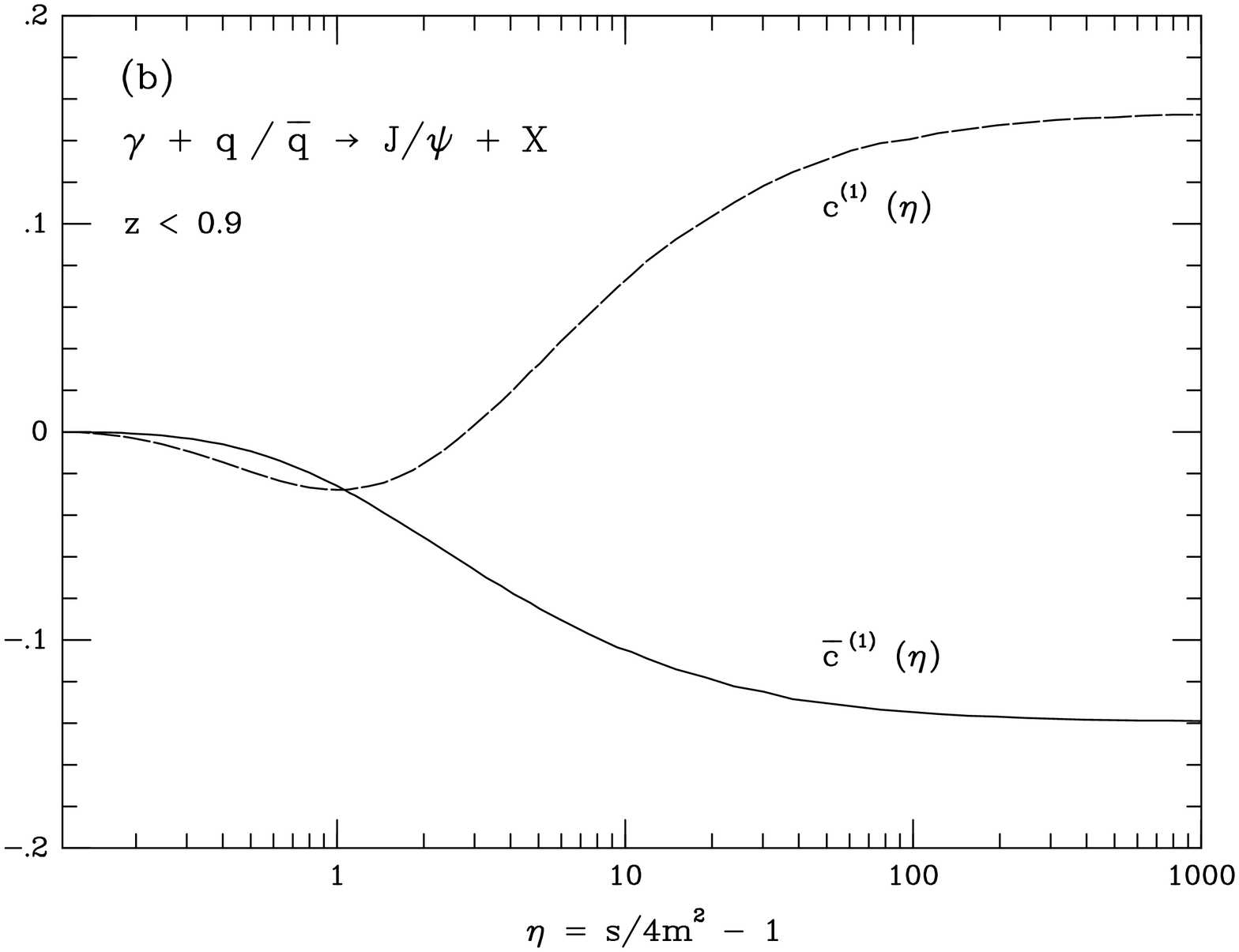,%
height=10cm,%
width=14.cm,%
bbllx=1.0cm,%
bblly=1.9cm,%
bburx=19.4cm,%
bbury=26.7cm,%
rheight=8.2cm,%
rwidth=15cm,%
angle=-90}

\vspace*{2.0cm}

\caption[xx]{  \label{f_scale}
                 (a) Coefficients of the QCD corrected total inelastic
                 [$z < 0.9$] cross section $\gamma + g \to  J/\psi + X$
                 in the physically relevant range of the scaling variable
                 $\eta = s_{\gamma p}/4m^2 - 1$; and (b) for $\gamma +
                 q/\overline{q} \to  J/\psi + X$.}

\vspace*{-5mm}

\end{figure}

The scaling functions $c_{\gamma i}(\eta)$ are shown in
Figs.\ref{f_scale}a/b for
the parton cross sections integrated over $z\le z_1$ where I have chosen
$z_1 = 0.9$ as discussed before.
[Note that the definition of $z$ is the
same at the nucleon and parton level since the momentum fraction $\xi$ of
the partons cancels in the ratio $z = {p\cdot k_\psi}\, / \,
{p\cdot k_\gamma}$.]
In the range $0.2{\,\,}\simlt{\,\,}\eta{\,\,}\simlt{\,\,}2$ the
hard gluon-radiation piece
$c_{\gamma g}^{(\mbox{\scriptsize{}1,H})}$ as well as
$\overline{c}_{\gamma g}^{(1)}$ differ from the curves in
Ref.\cite{kzsz94} by a few percent since the experimental cut $z < 0.9$
was not implemented properly in one term of \cite{kzsz94}.

The following comments can be inferred
from the figures. (i) The form of the scaling functions resembles the scaling
functions
in open-charm photoproduction \cite{sn92}. However, there is an important
difference. The "virtual + soft" contribution for $J/\psi$ production is
significantly more negative than for open-charm production. The destructive
interference with the lowest-order amplitude is not unplausible though,
as the momentum transfer of virtual gluons has a larger
chance [in a quasi-classical approach] to scatter quarks out of the small
phase-space element centered at $p_c +
p_{\overline{c}} = p_{J/\psi}$ than to scatter them from outside into this
small element.
(ii) While $c_{\gamma g}^{(0)}$ and
$c_{\gamma g}^{(\mbox{\scriptsize{}1,V+S})}$ scale asymptotically $\sim 1/s$,
the hard coefficients
$c_{\gamma g}^{(\mbox{\scriptsize{}1,H})}$ and
$c_{\gamma q}^{(1)}$ [as well as $\overline{c}_{\gamma g}^{(1)}$] approach
plateaus for high energies, built-up by the flavor excitation mechanism.
(iii) The cross sections on the quark targets are more than one order of
magnitude
smaller than those on the gluon target. (iv) A more detailed presentation
of the spectra would reveal
that the perturbative analysis is not under proper control in the limit $z\to
1$, as anticipated for this singular boundary region \cite{mk95}.
Outside the diffractive
region, i.e. in the truly inelastic domain, the perturbation theory is
well-behaved however.

\begin{figure}[hbtp]

\hspace*{0.75cm}
\epsfig{%
file=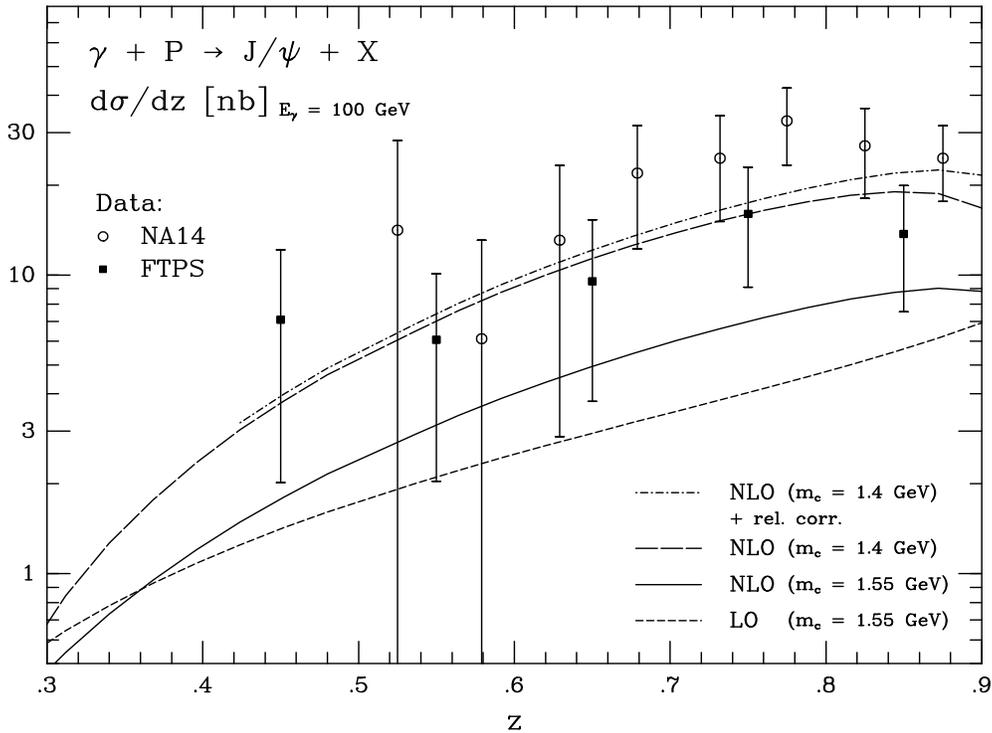,%
height=10cm,%
width=14.cm,%
bbllx=1.0cm,%
bblly=1.9cm,%
bburx=19.4cm,%
bbury=26.7cm,%
rheight=8.2cm,%
rwidth=15cm,%
angle=-90}

\vspace*{1.3cm}

\caption[xx]{
                 \label{f_zdistexp}
                 Energy spectrum $\md\sigma/\md{}z$,
                 at the initial photon energy $E_\gamma =
                 100$~GeV compared with the photoproduction
                 data \cite{na14,ftps}.}

\vspace*{-3mm}

\end{figure}

\begin{figure}[hbtp]

\hspace*{0.75cm}
\epsfig{%
file=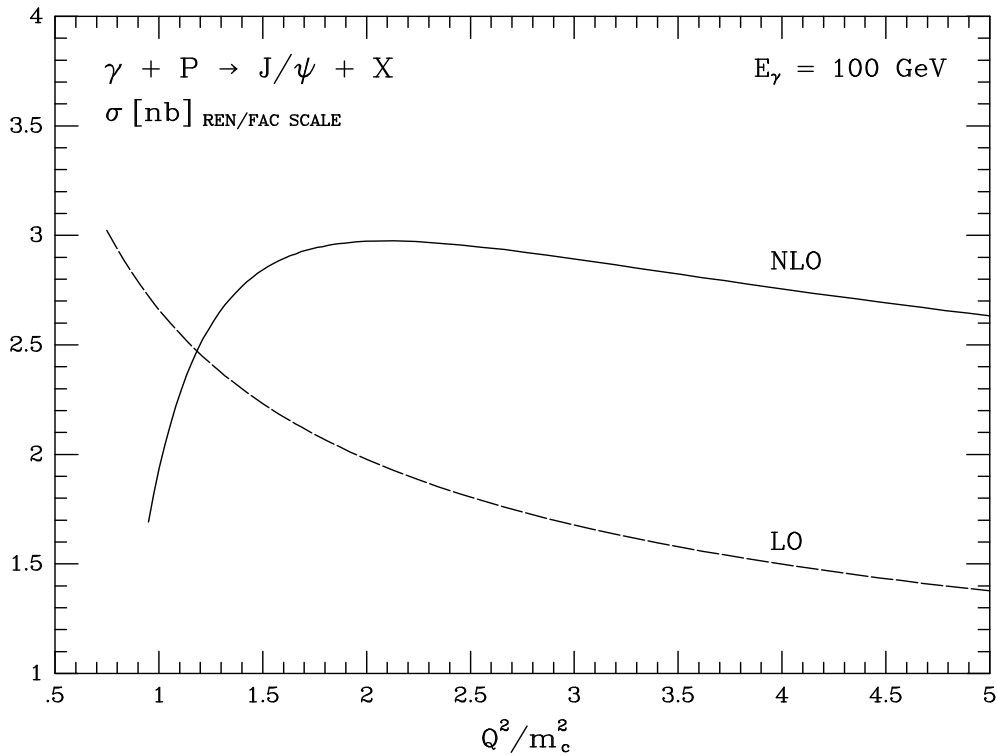,%
height=10.0cm,%
width=14.0cm,%
bbllx=1.0cm,%
bblly=1.9cm,%
bburx=19.4cm,%
bbury=26.7cm,%
rheight=8.2cm,%
rwidth=15cm,%
angle=-90}

\vspace*{1.5cm}

\caption[xx]{
            \label{f_qsqdep}
            Dependence of the total cross section
            $\gamma + P \to  J/\psi + X$ on the
            renormalization/factorization scale $Q$
            at a photon energy of $E_\gamma = 100$~GeV.}

\vspace*{-3mm}

\end{figure}

\begin{figure}[hbtp]

\hspace*{0.75cm}
\epsfig{%
file=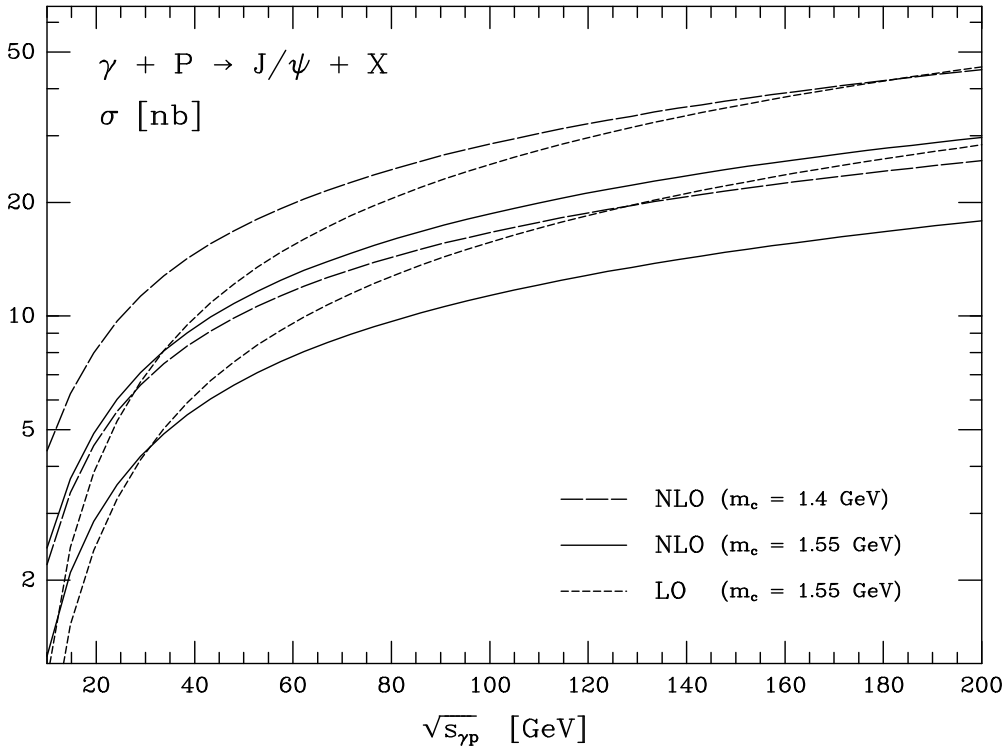,%
height=10.0cm,%
width=14.cm,%
bbllx=1.0cm,%
bblly=1.9cm,%
bburx=19.4cm,%
bbury=26.7cm,%
rheight=8.2cm,%
rwidth=15cm,%
angle=-90}

\vspace*{1.5cm}

\caption[xx]{
    \label{f_gptot}
    Total cross section for inelastic $J/\psi$ photoproduction
    $\gamma + P \to  J/\psi + X$ as a function of the
    photon-proton center of mass energy in the HERA energy range.}

\end{figure}

\begin{figure}[t]

\hspace*{0.95cm}
\epsfig{%
file=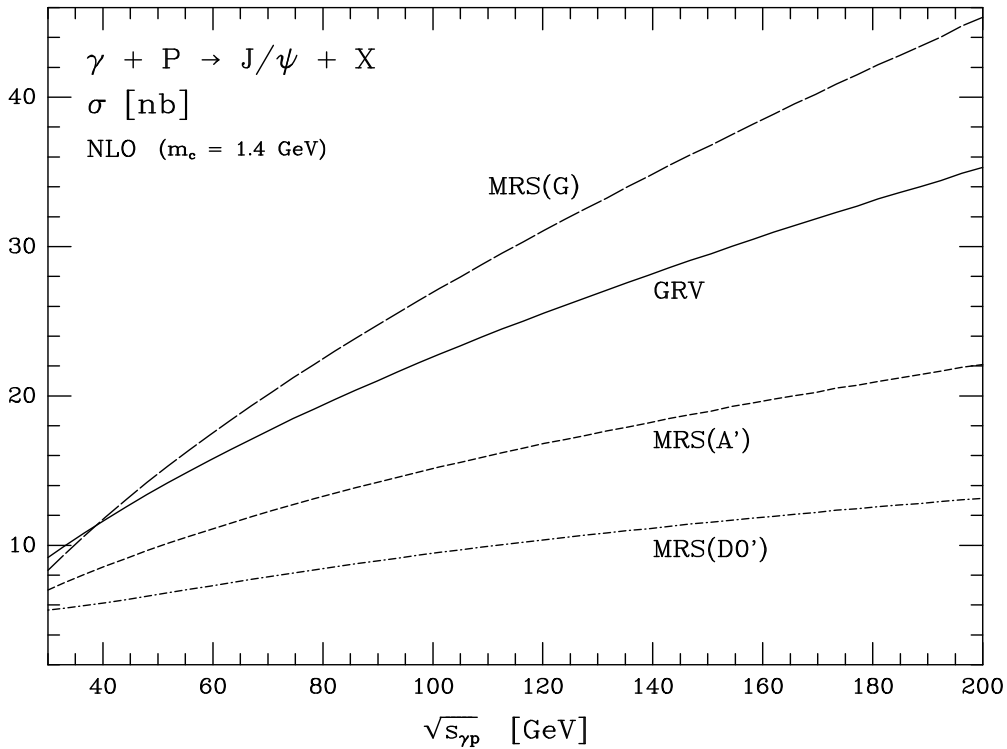,%
height=10.0cm,%
width=14.0cm,%
bbllx=1.0cm,%
bblly=1.9cm,%
bburx=19.4cm,%
bbury=26.7cm,%
rheight=8.2cm,%
rwidth=15cm,%
angle=-90}

\vspace*{1.6cm}

\caption[xx]{
    \label{f_gdist}
    The total cross section as a function of the
    photon-proton center of mass energy
    for different parametrizations of the gluon distribution
    of the proton.}
\end{figure}

The cross sections for $J/\psi$ photoproduction on nucleons are presented
in Figs.\ref{f_zdistexp}-\ref{f_gdist}.
In Fig.\ref{f_zdistexp} the leading-order and next-to-leading order
calculations are compared with the $J/\psi$ energy spectra of the two
fixed-target photoproduction experiments at photon energies near
$E_\gamma = 100$~GeV, corresponding to an invariant energy of
about $\sqrt{s\hphantom{tk}}\!\!\!\!\! _{\gamma p}\,\,  \approx 14$~GeV.
The GRV parametrizations of the parton densities \cite{grv94} have been
used. They are particularly suited to characterize the magnitude of the
radiative corrections properly since they allow one to compare the results
for the Born cross section folded with leading order parton densities, with
the cross sections consistently evaluated for parton cross sections
and parton densities in next-to-leading order. As the average momentum
fraction of the partons $<\! \xi\! > \sim 0.1$ is moderate, the curves are
not sensitive to the parametrization in the small-$x$ region. Similar to the
case of open heavy flavor production \cite{sn92,fmnr94}, the absolute
normalization of the cross section shows a strong dependence on the value
of the charm quark mass. In the static approximation the choice $m_c =
M_{J/\psi}/2$ is required for a consistent description of the heavy bound
state formation. However, a smaller mass value might be appropriate for a
reasonable description of the charm quark creation in the hard scattering
process. In order to demonstrate this uncertainty the results are shown for
two mass values, $m_c = M_{J/\psi}/2 \approx 1.55$~GeV and $m_c = 1.4$~GeV.
The $K$-factor,
$K=\sigma_{\mbox{\scriptsize{}NLO}}/\sigma_{\mbox{\scriptsize{}LO}} \sim 1.5$,
consists of two parts, one due to the QCD radiative corrections of the
leptonic $J/\psi$ width \cite{lepdecay} and a second part due to the
dynamical QCD corrections \cite{phd}, and does not strongly depend on $z$.
The dependence on the renormalization/factorization scale $Q$ is reduced
considerably in next-to-leading order. While the ratio of the cross sections
in leading order for $Q=m_c : (\sqrt{2}\,m_c) : M_{J/\psi}$ is given by
$1.8 : 1.3 : 1$, it is much closer to unity, $0.7 : 1.1 : 1$, in the
next-to-leading order calculation, Fig.\ref{f_qsqdep}.
The cross section runs through a maximum \cite{pms} near
$Q\approx \sqrt2{\,}m_c$ with broad width, the origin of the stable
behaviour in $Q$. In the BLM scheme \cite{blm} $Q$ moves from values below
$m_c$ at low energies up to $\sim \sqrt2{\,}m_c$ at the \mbox{HERA} energy
of $\sqrt{s\hphantom{tk}}\!\!\!\!\! _{\gamma p} \; \approx\; 100$~GeV.
In particular the value at high energies is significantly larger than
the corresponding BLM value for $J/\psi$ decays. The typical kinematical
energy scale is not set any more by the small gluon energy in the $J/\psi$
decay but rather by the typical initial-state parton energies.
I have adopted the scale $Q = \sqrt2{\,}m_c$ in Fig.\ref{f_zdistexp} and
subsequently.

In a systematic expansion one may finally add the relativistic corrections
as estimated in \cite{jkgw93}. Two conclusions can be drawn from the final
results presented in Fig.\ref{f_zdistexp}.
(i) The $J/\psi$ energy dependence d$\sigma/\mbox{d}z(\gamma + {\cal N}
\to J/\psi + X)$ is adequately accounted for by the color-singlet model
so that the shape of the gluon distribution in the nucleon can be extracted
from $J/\psi$ photoproduction data with confidence. (ii) The absolute
normalization  of the cross section is somewhat less certain; this is
apparent from the comparison with the photoproduction data. [The situation
is worse for electroproduction data \cite{emc,nmc}]. However, allowing for
higher-twist uncertainties of order $(\Lambda/m_c)^k \;\simlt\; 20\%$ for
$k \ge 1$, I conclude that the normalization too appears to be under
semi-quantitative control.

In Fig.\ref{f_gptot} I present the prediction of the cross section
for the \mbox{HERA} energy range, again for two values of the charm quark
mass, $m_c = M_{J/\psi}/2$ and $m_c = 1.4$~GeV, respectively.
In this high energy range the $K$-factor is smaller than at low energies,
$K=\sigma_{\mbox{\scriptsize{}NLO}}/\sigma_{\mbox{\scriptsize{}LO}} \sim 0.75$,
a consequence of the negative dip in the $c^{(1)}$ scaling function of
Fig.\ref{f_scale}. Note that the \mbox{LO} cross section in
Fig.\ref{f_gptot} has been evaluated by using leading-order expressions
for the parton distributions \cite{grv94}. When adopting the same set of
parton distributions for both \mbox{LO} and \mbox{NLO}
cross sections, the $K$-factor is close to one, depending in
detail on the photon-proton center-of-mass energy and the choice of the
parton distributions. The results in Fig.\ref{f_gptot} are shown for two
values of $\alpha_s(\sqrt2{\,}m_c) = 0.25$ and $0.31$ which correspond
to the $1\sigma$ lower and upper boundary of the error band in
Ref.\cite{pdg}, respectively. Since the cross section depends strongly on
the QCD coupling, I adopt this measured value, thus allowing for a slight
inconsistency to the extent that the GRV fits are based on a marginally lower
value of $\alpha_s$. For $m_c = M_{J/\psi}/2$ one finds,
for $z < 0.9$, a value of about $\sigma(\gamma + p \to J/\psi + X) \approx
18$~nb at an invariant $\gamma p$ energy of
$\sqrt{s\hphantom{tk}}\!\!\!\!\! _{\gamma p} \approx 100$~GeV; this
value rises to about 30~nb if one chooses $m_c = 1.4$~GeV and the larger
value 0.31 for the QCD coupling. Inclusion of the relativistic corrections
as estimated in Ref.\cite{jkgw93} increases the cross section in the
\mbox{HERA} energy range by approximately 10~\%.

Since the momentum fraction of the partons at \mbox{HERA} energies is
small, the cross section presented in Fig.\ref{f_gptot} is sensitive to
the parametrization of the gluon distribution in the small-$x$
region $<\! \xi\! > \sim 0.003$. This is demonstrated in Fig.\ref{f_gdist}.
The GRV parametrization adopted in Fig.\ref{f_gptot} leads to an almost
linear rise of the cross section with the $\gamma p$ c.m. energy. This
increase is even more pronounced when using the MRS(G) set and considerably
less marked for MRS(A') \cite{mrs95}. The MRSD0' set \cite{mrs} gives rise
to a much smaller cross section which does not strongly depend on the
$\gamma p$ energy in a wide range
$30$~GeV $< \sqrt{s\hphantom{tk}}\!\!\!\!\! _{\gamma p} \; < 200$~GeV.
Since the absolute normalization of the cross section is rather sensitive
to the value of $\alpha_s$ and the charm quark mass, the discrimination
between different parametrizations of the gluon density in the proton has
to rely on the shape of the cross section [as a function of the photon-proton
center of mass energy] rather than the absolute size of the prediction.

The cross sections for inelastic photoproduction of $\psi'$ particles can
be obtained from the results presented here by replacing the leptonic decay
width and multiplying with a phase space correction factor,
$ \sigma(\gamma P\to \psi'\;X) \approx \Gamma_{ee}^{\psi'} /
\Gamma_{ee}^{J/\psi}\, (M_{J/\psi} /M_{\psi'})^3
\times \sigma(\gamma P\to J/\psi\;X)
\approx$ \linebreak[4] $ 1/4 \times \sigma(\gamma P\to J/\psi\;X) $.
The production of $\Upsilon$ bottomonium bound states is suppressed, compared
with $J/\psi$ states, by a factor of about 300 at \mbox{HERA}, a consequence
of the smaller bottom electric charge and the phase space reduction by the
large $b$ mass.

\pagebreak[3]


\noindent
{\em Conclusion:} I have shown in this next-to-leading order perturbative
QCD analysis that the energy shape of the cross section for $J/\psi$
photoproduction is adequately described by the color-singlet model.
A semi-quantitative understanding has been achieved for
the absolute normalization of the cross section. Higher-twist effects
must be included to further improve the quality of the theoretical analysis.
The predictions for the \mbox{HERA} energy range provide a crucial test for
the underlying picture as developed so far in the perturbative QCD sector.

{\noindent\bf Acknowledgements}

\noindent
It is a pleasure to thank J.\ Zunft for a fruitful collaboration and
P.M.\ Zerwas for valuable discussions.
I have benefitted from conversations with W.J.\ Stirling and A.\ Vogt.


\end{document}